\documentclass[pre,twocolumn,floatfix,aps,showpacs]{revtex4}
\usepackage{amssymb}
\usepackage{graphicx}
\usepackage{epsf}
\usepackage{amsmath}
\usepackage{bm} %bold math
\usepackage{pspicture}
\usepackage{flafter}
\usepackage{float}
\begin{document}

\title{The universality of synchrony: critical behavior in a discrete
model of stochastic phase coupled oscillators}
\author{Kevin Wood$^{1,2}$, C. Van den Broeck$^{3}$, R. Kawai$^{4}$,
and Katja Lindenberg$^{1}$}
\affiliation{
$^{(1)}$Department of Chemistry and Biochemistry and Institute for
Nonlinear Science, and $^{(2)}$ Department of Physics,
University of California San Diego, 9500 Gilman Drive, 
La Jolla, CA 92093-0340, USA\\
$^{(3)}$Hasselt University, Diepenbeek, B-3590 Belgium\\
$^{(4)}$ Department of Physics, University of Alabama at Birmingham,
Birmingham, AL 35294 USA
}
\date{\today}

\begin{abstract}
We present the simplest discrete model to date that leads to
synchronization of stochastic phase-coupled oscillators.  In the mean
field limit, the model exhibits a Hopf bifurcation and global
oscillatory behavior as coupling crosses a critical value.  When
coupling between units is strictly local, the model undergoes a
continuous phase transition which we characterize numerically using
finite-size scaling analysis.  In particular, the onset of global
synchrony is marked by signatures of the XY universality class,
including the appropriate classical exponents $\beta$ and $\nu$, a lower
critical dimension $d_{lc} = 2$, and an upper critical dimension
$d_{uc}=4$. 
\end{abstract}


\maketitle

In the early 1960's, experiments with the Belousov-Zhabotinsky reaction
created a sensation by showing that dissipative structures
and self-organization in systems far from equilibrium correspond to real
observable physical phenomena. Since then, the breaking of time
translational symmetry has been a central theme
in the analysis of nonlinear nonequilibrium systems.  However, in the
later studies of spatially distributed systems, most of the interest
shifted to pattern forming instabilities, and surprisingly little
attention was devoted to the question of bulk oscillation and the
required spatial frequency and phase synchronization.  On the other
hand, the emergence of phase synchronization in populations of globally
coupled phase oscillators, with the synchronous firing of fireflies as
one of the spectacular examples, did generate intense
interest~\cite{sync}.  Because intrinsically
oscillating units with slightly different eigenfrequencies underlie the
macroscopic behavior of an extensive range of biological, chemical, and
physical systems, a great deal of literature
has focused on the mathematical principles governing the competition
between individual oscillatory tendencies and synchronous cooperation
\cite{kuramoto}.  While most studies have focused on
globally coupled units, leading to a mature understanding of the mean
field behavior of several models, relatively little work has examined
populations of oscillators in the locally coupled regime~\cite{local}.
The description of emergent synchrony has largely been
limited to small-scale and/or globally-coupled deterministic
systems~\cite{kuramoto}, despite the fact that the dynamics of
the physical systems in question likely reflect a combination of
finite-range forces and stochasticity.  Two recent studies by Risler
et al.~\cite{risler2} represent notable exceptions to this trend. 
They provide analytical evidence that locally-coupled identical noisy
oscillators belong to the XY universality class, though to date there
had been no empirical verification, numerical or otherwise, of their
predictions. 

\begin{figure}[b]
\begin{center}
\includegraphics[width=3.0 cm]{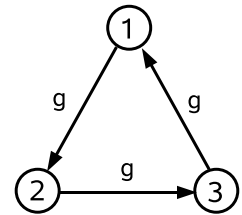}
\caption{Three state unit with generic transition rates $g$.}
\label{fig1}
\end{center}
\end{figure}

The difficulty with existing models of locally coupled oscillators is
that each is typically described by a nonlinear differential equation,
and it is notoriously computationally intensive to deal with systems of
coupled nonlinear differential equations,
especially if they also involve a stochastic component.
However, following Landau theory~\cite{goldenfeld}, macroscopically
observable changes occur without reference to microscopic specifics,
instead giving rise to classes of universal behavior whose members may
differ greatly at the microscopic level.  With this in mind, we
construct the simplest model with short-ranged interactions between
individual, stochastic, discrete phase
units exhibiting global phase synchrony and  amenable to extensive
numerical study.

Our starting point is a three-state unit~\cite{lutz} governed by
transition rates $g$ (Fig.~\ref{fig1}).  Loosely
speaking, we interpret the state designation as a generalized
(discrete) phase, and the transitions between states, which
we construct to be unidirectional, as a phase change and thus an
oscillation of sorts.  The probability of going from the current state
$i$ to state $i+1$ in an infinitesimal time $dt$ is $g dt$,
with $i$=1,2,3 modulo 3.  For an isolated unit, the transition rate is simply
a constant ($g$)
that sets the oscillator's intrinsic frequency; for many coupled units,
we will allow the transition rate to depend on the neighboring units in
the spatial grid, thereby coupling neighboring phases.  

For an isolated unit we write the linear evolution equation
$\partial P(t)/\partial t = M P(t)$, where the components $P_i(t)$ of the
column vector $P(t)= (P_1(t)~P_2(t)~P_3(t))^T$ 
are the probabilities of being in state $i$ at time $t$, and
\begin{equation}
M = \begin{pmatrix} -g & 0 & g \\ g & -g & 0 \\
0 & g & -g \end{pmatrix}.  
\label{Mmat}
\end{equation}
The system reaches a steady state for $P_1^*=P_2^*=P_3^*=1/3$.
The transitions $i\rightarrow i+1$
occur with a rough periodicity determined by $g$; that is, the time
evolution of our simple model qualitatively resembles that of the
discretized phase of a generic noisy oscillator.
\begin{figure}[bt]
\includegraphics[width = 8cm]{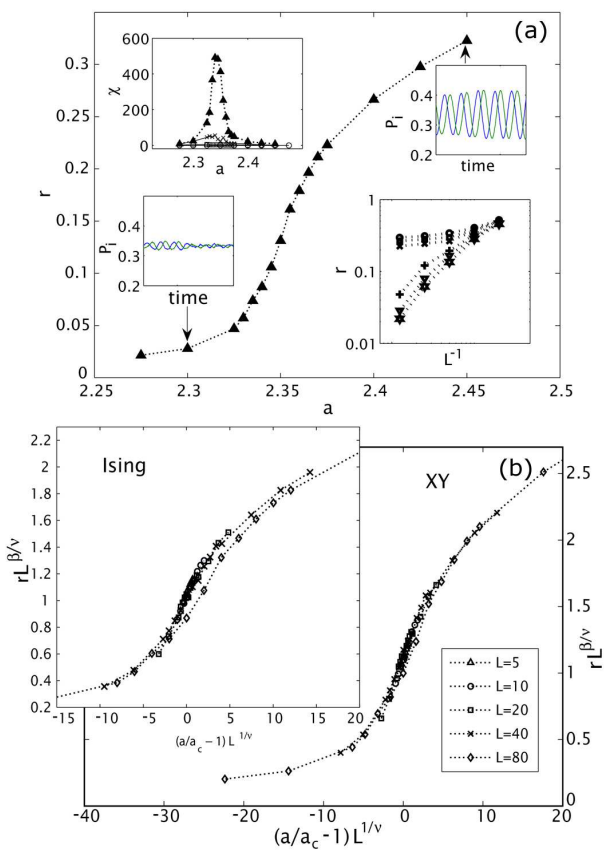}
\caption{({\bf a}) Onset of synchronization in $d$=3.
The system size $L$=80 is used.
Upper left inset: Fluctuations peak near the critical point, giving
an estimation of $a_c$=2.345$\pm$0.005. 
$P_1$ and $P_2$ undergo smooth temporal oscillations for
large $a$ (upper right inset), while lower $a$ decreases
temporal coherence (lower left inset). Lower right inset: Log-log plots of $r$
vs $L^{-1}$ with $a$=2.275, 2.3, 2.325, 2.375, 2.4, 2.425 from lowest to
highest plots.
({\bf b})
Finite size scaling analysis for $d$=3 using the XY and Ising critical
exponents. Data collapse with $a_c$=2.345.}
\label{fig2}
\end{figure}

The interesting behavior emerges when the transition probability
of a given unit to the state ahead depends on the
states of the unit's nearest neighbors in a spatial grid.  To capture
the physical nature of synchronization, we choose a function which
compares the phase at a given site with its neighbors, and adjusts the
phase at the given site so as to facilitate phase coherence.  With
universality in mind, we stress that the specific nature of the coupling
is not important so long as we ultimately observe a transition to global
synchrony at some finite value of the coupling parameter. 
For any unit, the transition rate from state $i$ to state $j$ is given by
\begin{equation}
g_{ij} = g \exp\left[{\frac{a(N_{j}-N_i)}{2
d}}\right]\delta_{j,i+1},
\label{gmu}
\end{equation}
where the constant $a$ is the coupling parameter
and $\delta$ is the Kronecker delta.
$N_k$ is the number of nearest neighbors in state $k$,
and $2d$ is the total number of nearest neighbors in $d$ dimensional cubic
lattices. 
While this choice is by no means unique and
these rates are somewhat distorted by their independence
of the number of nearest neighbors in state $i-1$, 
the form~(\ref{gmu}) is simplified by this
assumption and, as we shall see, does lead to synchronization.
  
To test for the emergence of global synchrony, we first consider
a mean field version of the model.
In the large $N$ limit with all-to-all coupling we write
\begin{equation}
g_{ij} = g \exp\left[a(P_{j} -
P_i)\right]\delta_{j,i+1} .
\label{gmuMF}
\end{equation}
Note that in the mean field limit 
$g_{ij}$ does not depend on the location of the unit within
the lattice.  Also, there is an inherent assumption that we can
replace $N_k/N$ with $P_k$.  With this simplification we arrive at a
nonlinear equation for the mean field probability,
$\partial P(t)/\partial t = M[P(t)]P(t)$,  with
\begin{equation}
M[P(t)] = \begin{pmatrix} -g_{12} & 0 & g_{31} \\ g_{12} & -g_{23}
& 0 \\ 0 & g_{23} & -g_{31} \end{pmatrix}.  
\label{Mmatmn}
\end{equation}

Normalization allows us to eliminate $P_3$
and obtain a closed set of equations for $P_1$
and $P_2$.  We can further characterize the mean field solutions
by linearizing about the fixed point $(P_1^*,P_2^*)=(1/3,1/3)$.
The complex conjugate eigenvalues of the Jacobian evaluated at the fixed
point, $\lambda_\pm = g(2a - 3 \pm i \sqrt{3})/2$,
cross the imaginary axis at $a=1.5$,
indicative of a Hopf bifurcation at this value, which following a
more detailed analysis~\cite{kuznetsov}
can be shown to be supercritical.
Hence, as $a$ increases, the mean field undergoes a qualitative change
from disorder
to global oscillations, and the desired global synchrony emerges.
Numerical solutions confirm this behavior,
yielding results that agree with simulations of an all-to-all coupling
array~\cite{ourlongerpaper}.  Here we characterize the breakdown of the
mean field description for the nearest-neighbor coupling model
as spatial dimension is decreased.

We perform simulations of the locally coupled model in
continuous time on $d$-dimensional cubic lattices with
periodic boundary conditions.  Time steps are 10 to
100 times smaller than the fastest local average transition
rate, i.e., $dt \ll e^{-a}$ (we set $g$=1). 
We find that much smaller time steps lead to essentially the
same results.  Starting from random initial conditions,
all simulations were run until an apparent steady state was reached, 
and statistics are based on 100 independent trials. 

\begin{figure}[bt]
\includegraphics[width = 8cm]{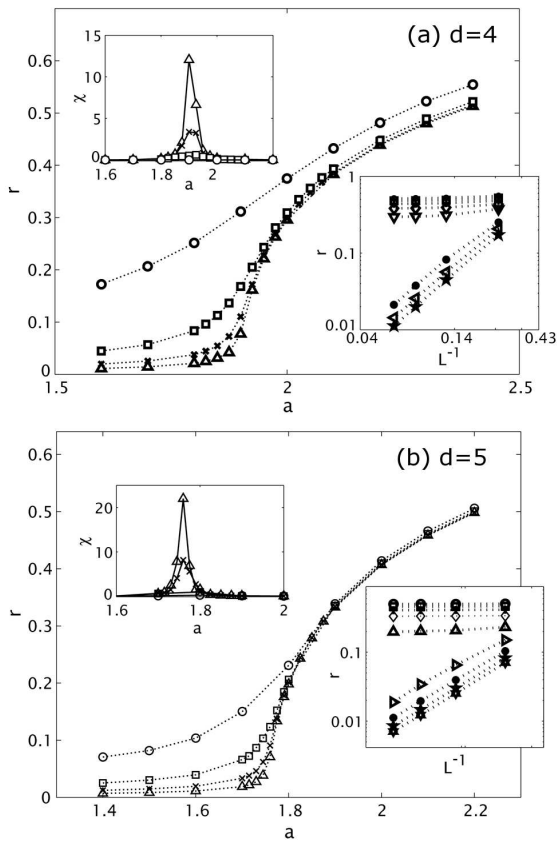}
\caption{Transition in $d$=4 (top) and $d$=5 (bottom): The order parameter
near the transition point is shown for various system sizes. 
$\bigcirc~L$=4, $\square~L$=8, $\times~L$=12, $\triangle~L$=16 for $d$=4
and $\bigcirc~L$=4, $\square~L$=6, $\times~L$=8, $\triangle~L$=10 for $d$=5.
The upper left inset in each panel shows the
generalized susceptibility which peaks
at $a$=1.900$\pm$0.025 for $d$=4 and at $a$=1.750$\pm$0.015 for $d$=5.
The lower right inset shows the system size
dependence of the order parameter.  For $d$=4 the coupling constant
varies from $a$=1.6 to 2.4 in increments of 0.1 (excluding $a$=1.9) from
lowest to highest plots and for $d$=5 from $a$=1.4 to 2.2 in increments of
0.1.}
\label{fig3}
\end{figure}

To characterize the emergence of phase synchrony, we introduce the
order parameter~\cite{kuramoto}
\begin{equation}
r=\langle R \rangle, \quad
R \equiv \frac{1}{N} \left \lvert \sum_{j=1}^N e^{i \phi_j} \right \rvert .
\end{equation}
Here $\phi_j$ is the discrete phase 2$\pi$($k$-1)/3
for state $k \in \lbrace 1,2,3 \rbrace$ at site $j$.
The brackets represent an
average over time in the steady state and over all
independent trials.  Nonzero $r$ in the thermodynamic limit indicates
synchrony.  We also calculate the generalized susceptibility
$\chi = L^{d} [ \langle R^2 \rangle - \langle R \rangle^2 ]$. 

In $d$=2 we do not see the emergence of global
oscillatory behavior.  Instead, we observe intermittent oscillations
(for very large values of $a$) that decrease drastically with
increasing system size. In fact, $r \rightarrow 0$ in the
thermodynamic limit, even for very large values of
$a$~\cite{ourlongerpaper}.  We conclude that the phase transition
to synchrony cannot occur for $d$=2. Interestingly, snapshots of the
system reveal increased spatial clustering as $a$ is increased, as well
as the presence of defect structures, perhaps indicative of
Kosterlitz-Thouless-type phenomena~\cite{goldenfeld}.
Further studies along these lines are underway.

\begin{figure}[t]
\includegraphics[width = 8cm]{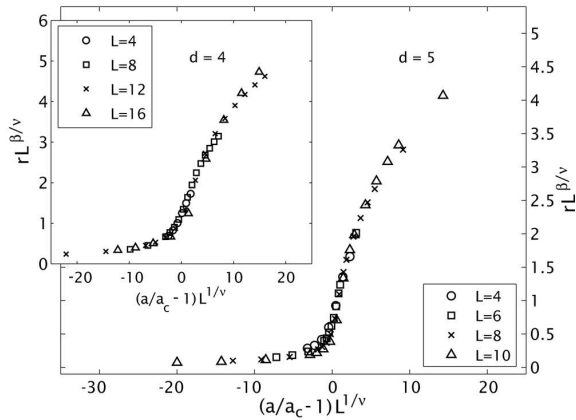}
\caption{Finite size scaling analysis for $d$=4 and $d$=5: Data collapse using 
ansatz (\ref{fss}) with mean field exponents.}
\label{fig4}
\end{figure}

In contrast to the $d$=2 case, which serves as the lower critical
dimension, a clear thermodynamic-like phase transition occurs in three
dimensions.  We see the emergence of global oscillatory behavior
as $a$ increases past a critical value $a_c$. 
Figure~\ref{fig2}a shows the behavior of the order parameter as
$a$ is increased for the largest system studied ($L$=80); the upper left
inset shows the peak in $\chi$ at $a$=2.345$\pm$0.005, thus
providing an estimate of the critical point $a_c$.
We see no change as system size is increased beyond
$L$=40.  At any rate, finite size effects are within the range of
our estimation. 
The lower right inset in Fig.~\ref{fig2}a shows explicitly that
for $a < a_c$, $r \rightarrow 0$
as system size is increased, and a disordered phase
persists in the thermodynamic limit. For $a>a_c$, the order parameter
approaches a finite value as the system size increases. 
We tried to apply the Binder cumulant crossing
method~\cite{binder} for determining $a_c$ more precisely, but residual
finite size effects and statistical uncertainties in the data prevent
us from determining the crossing point with more precision than that
stated above.  In any case, the accuracy of our current estimation of
the critical point suffices
to determine the universality class of the transition. 

To further characterize this transition, we use 
finite size scaling analysis
by assuming the standard scaling 
\begin{equation}
r = L^{-\frac{\beta}{\nu}} F[(a-a_c) L^\frac{1}{\nu} ].
\label{fss}
\end{equation}
Here $F(x)$ is a scaling function that approaches a constant
as $x \rightarrow 0$. 
To test our numerical data against different universality classes we
choose the appropriate critical exponents for each, recognizing that
there are variations in the reported values of these
exponents~\cite{pelissetto}. For the
XY universality class we use the exponents
$\beta$=0.34 and $\nu$=0.66~\cite{gottlob}. 
For the Ising exponents we use
$\beta$=0.31 and $\nu$=0.64~\cite{huang}.
In Fig.~\ref{fig2}b, we see quite convincingly a collapse
when exponents from the XY class are used.  For comparison, we also show
the data collapse with 3D Ising exponents (note the scale differences).

For $d$=4 we estimate the transition coupling to be $a_c$=1.900$\pm$0.025 from
the peak in $\chi$ (Fig.~\ref{fig3}a). 
Because we expect $d$=4 to be the upper critical dimension in accordance
with XY/Ising behavior, we anticipate a slight breakdown of the
scaling relation (\ref{fss}).  A priori it is not
clear how strongly (\ref{fss}) will be violated in $d$=4.
As shown in Fig.~\ref{fig4}, the data collapse is very good
with the mean field exponents.
As such, our simulations suggest that $d$=4 serves as the upper
critical dimension; additionally, it appears that corrections to
finite-size scaling at $d$=4 are not substantial, though a much more
precise study would be needed to investigate such corrections in greater
detail.  

To further support the claim that $d_{uc}$=4, we consider the
case $d$=5. We see a transition to synchrony 
at $a_c$=1.750$\pm$0.015 (Fig.~\ref{fig3}b).  As expected, this
value is considerably closer than the critical coupling in
four dimensions to the mean field value $a_c$=1.5.
The data collapse with the mean field exponents is
excellent, as shown in Fig.~\ref{fig4}.  We note the
rarity of computations in such a high dimension.

In conclusion,
while nonequilibrium phase transitions exhibit a much
wider diversity in universality classes than equilibrium ones~\cite{net},
it is remarkable that the prototype of a nonequilibrium transition,
namely, a phase transition that breaks the symmetry of translation in
time, is described by an equilibrium universality class.  In particular,
the Mermin-Wagner theorem, stating that continuous symmetries can not
be broken in dimension two or lower, appears to apply. The
XY model is known to display a Kosterlitz-Thouless transition in which,
beyond a critical temperature, vortex pairs can unbind into individual
units creating long range correlations. Preliminary results indicate
that a similar transition occurs in our model. 

Finally, a note of
caution concerning the discreteness of the phase is in order. We
first note that microscopic models often feature discrete degrees of
freedom. For example, our model is reminiscent of the triangular reaction
model of Onsager~\cite{onsager}, on the basis of which
he illustrated the concept of detailed balance as a characterization
of equilibrium. Continuous phase models appear in a suitable
thermodynamic limit. We stress that the breaking of time translational
symmetry can occur independently of whether the phase is a discrete or
continuous variable. 
It is, however, not evident whether continuous
and discrete phase models belong to the same universality class.
For example, the three state ferromagnetic
Potts model displays a weak first order
phase transition in $d$=3~\cite{potts}, while the anti-ferromagnetic version
belongs to the
XY universality class~\cite{pelissetto,antipotts}. The
results found here appear to be compatible with the latter,
but a renormalization calculation confirming this hypothesis would
be welcome.

\section*{Acknowledgments}

This work was partially supported by the National Science Foundation
under Grant No. PHY-0354937.

\end{document}